\documentclass[apjl]{emulateapj}
\usepackage[dvips]{color}
\usepackage{times}

\shorttitle{}
\shortauthors{Bischoff-Kim}

\begin{document}

\title{Asteroseismology of the Kepler field DBV White Dwarf \-- It's a hot one!}

\author{Agn\`{e}s Bischoff-Kim}  
\affil{Chemistry, Physics and Astronomy Department, Georgia College \& State University,
    Milledgeville, GA 31061; \textcolor{blue}{agnes.kim@gcsu.edu}}

\and

\author{Roy H. \O stensen}
\affil{Instituut voor Sterrenkunde, K.U. Leuven, Celestijnenlaan 200D, B-3001 Leuven; \textcolor{blue}{roy@ster.kuleuven.be}}

\begin{abstract}

We present an asteroseismic analysis of the helium atmosphere white dwarf (a DBV) recently found in the field of view of the \emph{Kepler} satellite. We analyze the 5-mode pulsation spectrum that was produced based on one month of high cadence Kepler data. The pulsational characteristics of the star and the asteroseismic analysis strongly suggest that the star is hotter (29200~K) than the 24900~K suggested by model fits to the low S/N survey spectrum of the object. This result has profound and exciting implications for tests of the Standard Model of particle physics. Hot DBVs are expected to lose over half of their energy through the emission of plasmon neutrinos. Continuous monitoring of the star with the Kepler satellite over the course of 3 to 5 years is not only very likely to yield more modes to help constrain the asteroseismic fits, but also allow us to obtain a rate of change of any stable mode and therefore measure the emission of plasmon neutrinos.

\end{abstract}

\keywords{Dense matter --- Elementary particles --- Stars: oscillations --- Stars: variables: general --- white dwarfs}

\section{Introduction}

Much to everyone's surprise considering the statistical odds, a helium atmosphere pulsating white dwarf (a DBV) was discovered in the 105 square degree part of the sky monitored by the \emph{Kepler} satellite. GALEXJ192904.6+444708 (a.k.a WD\ J1929+4447 or KIC 8626021 in the \emph{Kepler} input catalog) was found to be a good DBV candidate in a small auxillary survey undertaken by \O estensen et al. (2011a), following the null-detection of pulsations in the 17 WDs observed during the survey phase of the {\em Kepler} mission (\O estensen et al.~2010, 2011b). A low S/N survey spectrum of the object revealed its physical parameters to be $T_{\rm eff}=24900 \pm 750$ K and log $g=7.91 \pm 0.07$, placing it right in the middle of the DBV instability strip. One month of short cadence (58.8\ s exposures) data was collected by the \emph{Kepler} spacecraft in October and November 2010, revealing a pulsation spectrum with five roughly equally spaced pulsation modes, three of which appears as evenly split triplets. The Fourier Transform and pulsational properties are presented in the discovery paper (\O stensen et al.~2011a). In this paper, we present the first asteroseismic analysis of WD\ J1929+4447.

Our asteroseismic analysis strongly suggests that WD\ J1929+4447 is more likely a blue edge DBV, with a pulsation spectrum reminiscent of EC\ 20058-5234's (a.k.a QU Tel, Koen et al.~1995). This makes the star even more interesting as it then becomes the second known hot DBV with stable periods. Stable hot DBVs are good candidates to study the emission of plasmon neutrinos (Winget et al.~2004). An observed rate of change of the periods of any phase-stable mode, combined with models, will allow us to place constraints on the emission of plasmon neutrinos. This concept was attempted with EC\ 20058-5234 (Dalessio et al.~2010; Bischoff-Kim 2008), but without much success because of crowding of the field around the star. WD\ J1929+4447 does not suffer from such pollution and may yield more positive results. Evolutionary rates of period changes have succesfully been measured for other pulsating white dwarfs, the most notable example being G117-B15A (e.g. Kepler et al.~2005). Hot DBVs cool a factor of a hundred times faster than a cooler white dwarf like G117-B15A (C\'orsico \& Althaus 2004). If it were not for the pollution from a field star for EC\ 20058-5234, an evolutionary rate of period change could have been obtained for that object based on a 5 year baseline of data.

In section \ref{observations} we summarize the observed pulsational properties of WD\ J1929+4447. In section \ref{fits} we use these properties to perform an asteroseismic study of that star and find a temperature of 29200\ K, at odds with the spectroscopy. We discuss this discrepancy and other results in section \ref{discussion}. We summarize and conclude in section \ref{conclusions}.

\section{Pulsational properties}
\label{observations}

A total of 5 independent modes were found in the high signal-to-noise Fourier Transform (FT) of the data collected by the \emph{Kepler} satellite. The 5 modes are listed in table \ref{t1}). We reproduced and preserved the labels given to each mode (first column in table \ref{t1} from \citealt{Ostensen11}) for consistency. The modes labeled $f_{1,0}$, $f_{2,0}$, and $f_{3,0}$ are the central members of clearly defined triplets in the FT. This allows us to positively identify these modes as $\ell=1$, $m=0$ modes and helps us constrain the asteroseismic fits. $\ell=1$ triplets are caused by rotation of the star. From the frequency splitting of these triplets, \citeauthor{Ostensen11} determine a rotation period of 1.2 day for the star, typical for white dwarfs. Table \ref{t1} also shows the periods for our best fit model, discussed in section \ref{fits}.

\begin{table}
\begin{center}
\caption{
\label{t1}
Observed periods and the corresponding periods for the best fit model
}
\begin{tabular}{lccclr}
\tableline\tableline
Mode       & Observed     & Amplitude & Model      & $\ell$             & $k$ \\
Label      & Period (s)   & (mma)     & Period (s) &                    &     \\
\tableline
$f_{2,0}$  & 197.11       & 2.92      & 197.28     & 1\tablenotemark{a} & 3   \\
$f_{1,0}$  & 232.02       & 5.21      & 232.46     & 1\tablenotemark{a} & 4   \\
$f_{3,0}$  & 271.60       & 1.87      & 271.27     & 1\tablenotemark{a} & 5   \\
$f_4    $  & 303.56       & 1.29      & 303.64     & 1                  & 6   \\
$f_5    $  & 376.10       & 1.05      & 375.87     & 2                  & 15  \\
\tableline
\end{tabular}
\tablenotetext{1}{These modes are identified as $\ell$\,=\,1, $m$\,=\,0 based on being the central members of well-defined triplets in the FT.}
\end{center}
\end{table}

As noted in the discovery paper, WD J1929+4447 is pulsationally most similar to EC\ 20058-5234, a blue edge DBV. Both stars have short period modes observed in their pulsation spectrum. EC\ 20058-5234 has at least 6 independent modes under 400 seconds and 2 around 500 seconds (Sullivan et al.~2008). But EC\ 20058-5234 has a temperature spectroscopically and asteroseismically determined to be around 28000\ K (Beauchamp et al.~1999; Bischoff-Kim \& Metcalfe 2011), while \O stensen et al. (2011a) determined a temperature of 24900\ K for WD\ J1929+4447, placing it in the middle of the DBV instability strip and not on the blue edge.

\section{Asteroseismic analysis}
\label{analysis}

\subsection{The Models}
\label{models}

To compute all our models, we used the White Dwarf Evolution Code (WDEC). The WDEC evolves hot polytrope models from temperatures close to 100,000K down to the temperature of our choice. Models in the temperature range of interest for the present study are thermally relaxed solutions to the stellar structure equations. Each model we compute for our grids is the result of such an evolutionary sequence. 

The WDEC is described in detail in Lamb \& van Horn (1975) and Wood (1990). We used smoother core composition profiles and experimented with the more complex profiles that result from stellar evolution calculations (Salaris et al.~1997). We updated the envelope equation of state tables from those calculated by Fontaine et al. (1977) to those given by Saumon et al. (1995). We use OPAL opacities (Iglesias \& Rogers 1996) and plasmon neutrino rates published by Itoh et al. (1996). 

\begin{figure}
\includegraphics[width=\hsize,clip]{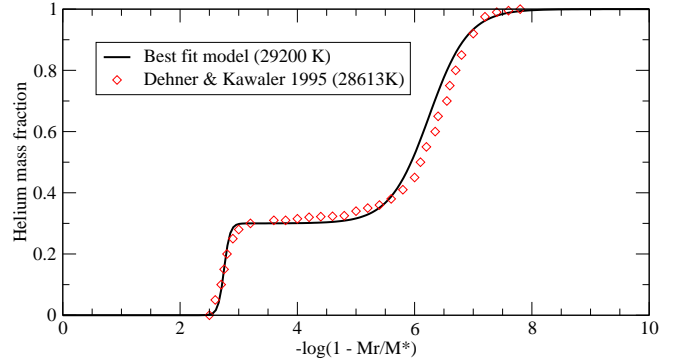}
  \caption{
  \label{f1}
  Helium mass fraction of model DBs. The symbols show the profile that results from the time dependent diffusion calculations of  Dehner \& Kawaler (1995). The solid line shows the profile for our best asteroseismic fit of WD\ J1929+4447 (see section \ref{fits}).
  }
\end{figure}

DBVs are younger than their cooler cousins the DAVs. Time dependent diffusion calculations (e.g. Dehner \& Kawaler 1995) show that at 24000\ K, a typical temperature for a DBV, the carbon has not fully settled into the core of the star yet. We expect the helium layer to be separated into a mixed He/C layer with a pure He layer on top, as shown in Fig. \ref{f1}. Following  Metcalfe (2005), we adopted and parameterized this structure in our models. 

We calculated grids of models using the WDEC and then ran a fitting subroutine to match the periods of the models ($\rm P_{calc}$) with the observed periods and calculated residuals using the usual formula

\begin{equation}
\label{eq1}
\rm \sigma_{RMS} = \sqrt{\frac{\sum_{1}^{n_{obs}} {(P_{calc}-P_{obs})^2}}{n_{obs}}},
\end{equation}

\noindent where $\rm n_{obs}$ is the number of periods present in the pulsation spectrum.

\subsection{Parameters and grids}
\label{grids}

In our asteroseismic fits, we vary up to 6 parameters; the effective temperature, the mass and 4 structure parameters. There are two parameters associated with the shapes of the oxygen (and carbon) core composition profiles: the central oxygen abundance ($\rm X_o$), and the edge of the homogeneous carbon and oxygen core ($\rm q_{fm}$). For envelope structure, $\rm M_{env}$ marks the location of the base of the helium layer and $\rm M_{He}$ the location where the helium abundance rises to $1$ (see Fig. \ref{f1}). $\rm M_{env}$ and $\rm M_{He}$ are mass coordinates, defined as e.g. $\rm M_{env} = -\log(1 - M(r)/M_*)$, where M(r) is the mass enclosed in radius r and $\rm M_*$ is the stellar mass. 

With only 5 periods, 6 parameter fits are underconstrained. While there is much to learn from attempting 6 parameter fits, we settled for 4 parameters (later relaxing a $5^{th}$). There are three logical ways of fixing 2 of the 6 parameters: 1) fix the mass and effective temperature according to the spectroscopy, 2) fix the core structure parameters, and 3) fix the envelope parameters. In light of the basic pulsational properties of the object (see also section \ref{periodspacing}), option 1) seems the least promising. Of the other two choices, we decided to first try fixing the envelope structure to that of Dehner \& Kawaler (1995). This meant that $\rm M_{env}$ was fixed to -2.80 for all models and $\rm M_{He}$ was first fixed to -5.50 for models between 20000\ K and 22000\ K, -5.90 for models between 22200\ K and 26000\ K and -6.30 for models between 26200\ K and 28000\ K, to match the profiles resulting from the diffusive settling computed by Dehner \& Kawaler (1995).

We calculated 2 grids of models; one covering a broad range of parameter space at lower resolution and one zooming in on the promising area of parameter space with higher resolution. The range of parameters and step sizes for each are summarized in table \ref{t2}. The broad grid contains 305,820 models that succesfully converged (out of 315,700 total computed) and the higher resolution grid contains 403,065 converged models, with a similar success rate. Table \ref{t2} also shows the parameters for the best fit models, discussed in section \ref{fits}.

\begin{table*}
\begin{center}
\caption{
\label{t2}
Parameters covered in the model grids and best fit parameters for WD\ J1929+4447
}
\begin{tabular}{c|ccc|ccc|c}
\tableline\tableline
Parameter         & \multicolumn{3}{c|}{Grid 1}          & \multicolumn{3}{c|}{Grid 2}                        & Best fit         \\
                 & Minimum     & Maximum  & Step Size  & Minimum          & Maximum        & Step Size     &                  \\
\tableline
$\rm T_{eff}$ (K) & 20000     & 28000  & 200        & 24100          & 30000        & 100           & 29200            \\
$\rm M_*$ (Solar) & 0.500       & 0.690    & 0.010      & 0.555            & 0.650          & 0.05          & 0.570            \\
$\rm M_{env}$     & \multicolumn{3}{c|}{-2.80 (fixed)}   & \multicolumn{3}{|c|}{-2.80 (fixed)}              & -2.80            \\
$\rm M_{He}$      & \multicolumn{3}{c|}{-5.50, -5.90, or -6.30 (see text)} & -6.10 & -6.50  & 0.20          & -6.30            \\
$\rm X_o$         & 0.00        & 1.00     & 0.10       & 0.40             & 1.00           & 0.05          & 0.60, 0.65\tablenotemark{a} \\
$\rm q_{fm}$ ($\rm M_*$) & 0.10 & 0.80     & 0.02       & 0.32             & 0.48           & 0.02          & 0.36             \\
\tableline
\end{tabular}
\tablenotetext{1}{Close tie}
\end{center}
\end{table*}

\subsection{Average period spacing}
\label{periodspacing}

The average period spacing provides a useful asteroseismic measure of the mass and temperature of the star, independent of the details of internal chemical composition profiles. Higher $k$ modes are not trapped in the core and according to asymptotic theory, they are evenly spaced in period. This spacing is given by (Unno et al.~1989)

\begin{equation}
\label{eq2}
\Delta P = \frac{\pi}{\ell(\ell + 1)^{1/2}}\left[\int_{r_1}^{r_2}\frac{N}{r}dr\right]^{-1},
\end{equation}

\noindent
where $r_1$ and $r_2$ are turning points of the mode and N is the Brunt-V\"{a}is\"{a}l\"{a} frequency. The asymptotic period spacing is $\ell$ dependent, with higher $\ell$ modes having smaller spacing. The dependence on the Brunt-V\"{a}is\"{a}l\"{a} frequency leads to higher mass and higher temperature models having a smaller period spacing. 

In the case of WD\ J1929+4447, we must be careful when using average period spacing arguments. First, the modes are not higher $k$ modes, so at least a subset of them may be strongly trapped in the core (we are not in the asymptotic limit). Second, we only have a few modes to give us an average period spacing. That being said and somewhat ignoring this warning, we proceed.

From the first 4 modes in table \ref{t1} we determine an average period spacing of 35.9 seconds. This is to compare to 36.5 seconds for EC\ 20058-5234 ($\sim 28000$\ K) and 38.8 seconds for GD358 ($\sim 25000$\ K, \citealt{Provencal09}). The former marks the current blue edge of the DBV instability strip, while the latter is the prototype of a middle of the instability strip DBV. From these numbers alone we might conclude that WD\ J1929+4447 is closer in temperature (and mass) to EC\ 20058-5234 than it is to GD358, at odds with the spectroscopy.

\begin{figure}
  \includegraphics[width=\hsize,clip]{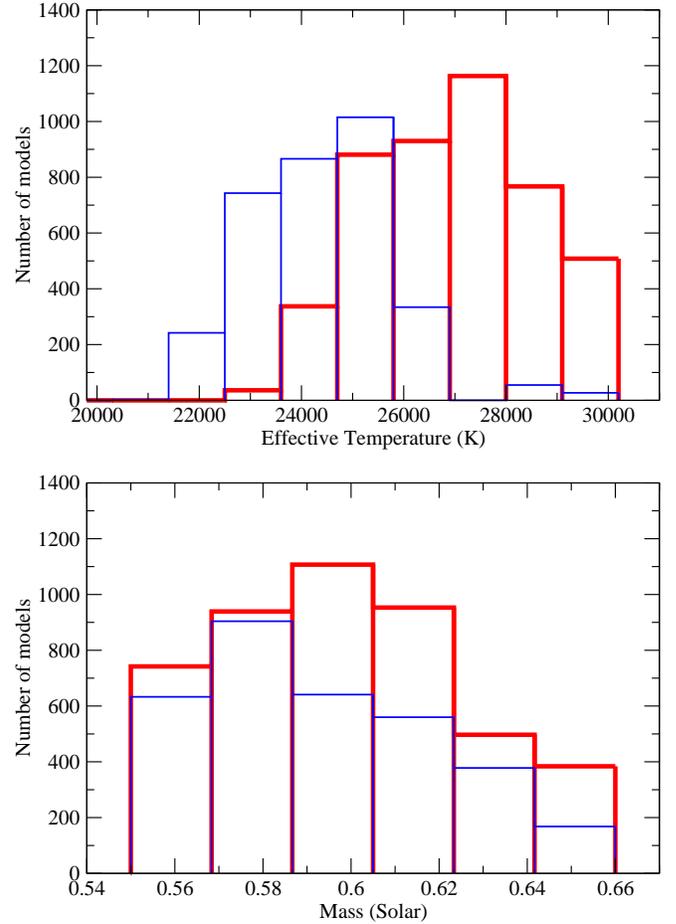}	
  \caption{
  \label{f2}
  Temperature (upper panel) and mass (lower panel) distribution of a subset of model DBs. The thick red lined histograms show the distribution for models that have an asymptotic period spacing between 36 and 37 seconds (like EC\ 20058-5234) and the thin blue lined histograms show the distribution for asymptotic period spacings between 38.2 and 39.2 seconds (like GD358).
  }
\end{figure}

To gain a better notion of how a 3 second difference in average period spacing translates into a difference in mass and effective temperature, we calculated asymptotic period spacings for the models in our grids and pulled models with period spacings similar to that of EC\ 20058-5234 and another set with spacings similar to that of GD358 (performing some normalizing in resolution to merge grid 1 and grid 2 - 31500 models total). We show the effective temperature and mass distribution of the two populations in Fig. \ref{f2}. The 36.5 second period spacing population peaks at an effective temperature of 27500\ K while the other peaks at 25500\ K, consistent with the effective temperatures of the two stars. This suggests that WD\ J1929+4447, with its shorter period spacing, is hotter than 24900\ K.

\begin{figure*}
\center
  \includegraphics[width=0.75\hsize,clip]{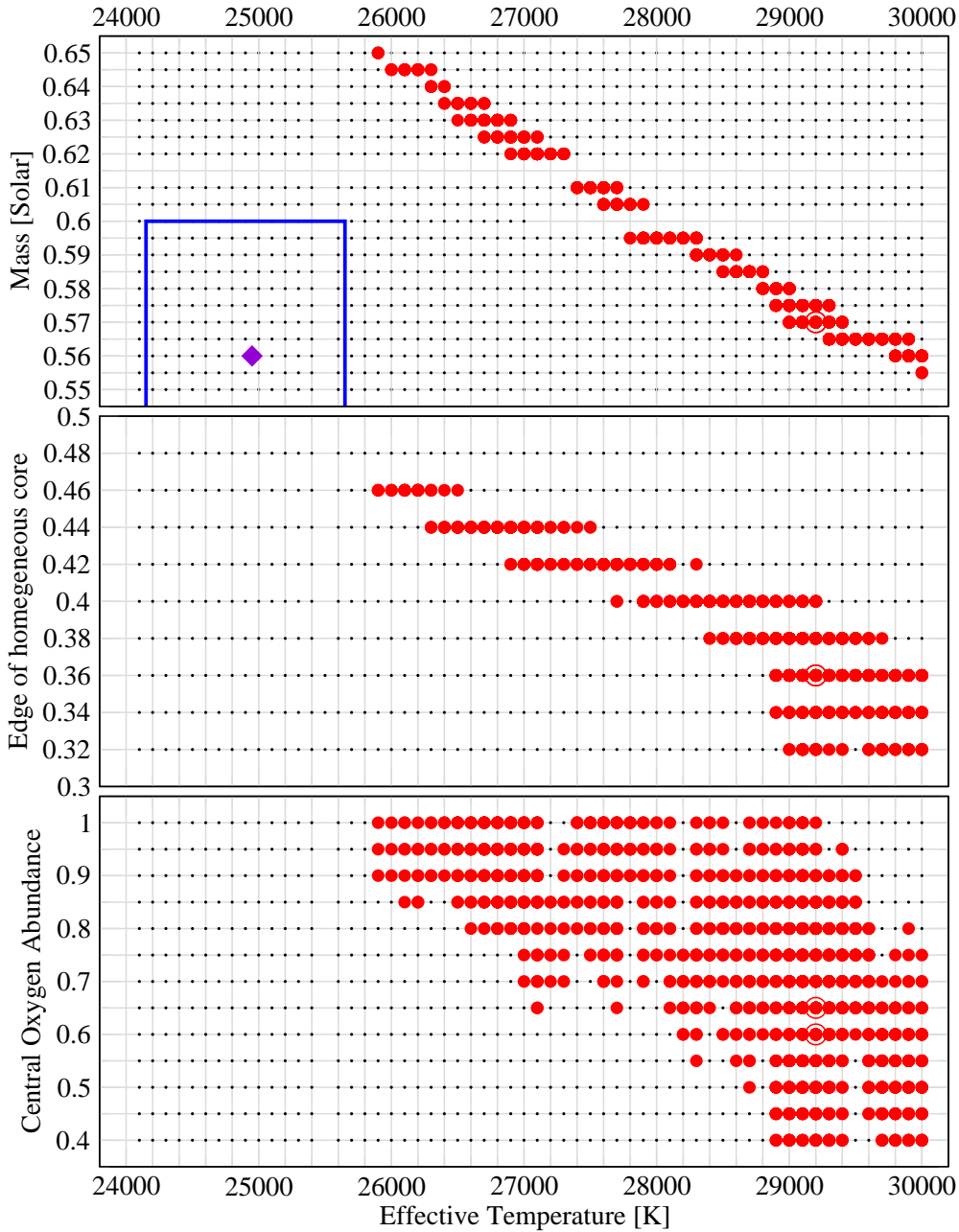}	
  \caption{
  \label{f3}
  Best fit models in 3 different planes of parameter space. The filled circles are models that fit the observed period spectrum with $\rm \sigma_{RMS} \leq 1.0 \; s$. The best fit models at 29200\ K is circled. In the top panel (mass versus effective temperature), we also mark the spectroscopic value (diamond) and the formal 1\--$\sigma$ fitting error region around it.}
\end{figure*}

\subsection{Results of the asteroseismic fitting}
\label{fits}

The parameters of the two best fit models that resulted from our grid search are listed in table \ref{t2} and the periods of the marginally better one ($\rm X_o = 0.60$) in table \ref{t1}. For these models,  $\rm \sigma_{RMS} = 0.28 \; s$. To get a sense of the quality of this fit, it is useful to compare it to similar fits performed in the litterature. The difficulty in comparing the results of the asteroseismic analysis of different pulsating white dwarfs is that the number of observed periods to fit and the number of parameters used varies. One way to compare them is to use the BIC parameter (Bayes Information Criterion). The BIC parameter measures an absolute quality of the fit, by taking into account differing numbers of data points and free parameters \citep{Koen00}. It penalizes fits involving a greater number of parameters relative to the number of data points. Applied to asteroseismology, it is given by

\begin{equation}
\rm BIC = n_{par}\frac{\ln(n_{obs})}{n_{obs}} + n_{obs} \ln(\sigma_{RMS}^2),
\end{equation}

\noindent where $\rm n_{par}$ is the number of free parameters in the fit. A smaller BIC parameter indicates a better fit.

For the present case, $\rm n_{obs} = 5$ and $\rm n_{par} = 5$. With $\rm \sigma_{RMS} = 0.28 \; s$, this translates to $\rm BIC = -0.41$. Among the best recent asteroseismic fits (Bischoff-Kim et al.~2008; Castanheira \& Kepler 2008), using the BIC, only one fit among 6 comes out superior. But that is a fit to a single mode star using 4 parameters. It is difficult to argue for the significance of such a fit.

It is illuminating to also look at the parameters (mass and effective temperature) of models that did not fit quite as well, but were still good fits. Fig. \ref{f3} is a plot of where all the models that fit to better than $\rm \sigma_{RMS} = 1.0 \; s$ lie in parameter space. The best fit models follow a negatively sloping trend in the mass versus effective temperature plane. This is a direct consequence of Eq. \ref{eq2}. To fit a fixed set of observed periods, higher temperature models must compensate with a lower mass in order to keep the appropriate average period spacing. This trend frequently shows up in white dwarf asteroseismology (e.g. Bischoff-Kim et al.~2008; Castanheira \& Kepler 2008). The band of acceptable model fits in that plane misses the 1-$\sigma$ spectroscopic box completely. However, we do not consider this a failure as it is well known that the He I lines in the DB instability region change only weakly with temperature. Thus, small calibration effects can easily throw the temperature by many times the formal 1-sigma fitting error. The mass of our best fit models, on the other hand, corresponds well with the mass determined from the spectroscopic surface gravity.

Here it is worth summarizing our method and emphasizing one aspect of it before presenting one more result. We started out by fixing the helium envelope parameters to the values that fit the time dependent diffusion calculations of Dehner \& Kawaler (1995) but when we zoomed in on the promising part of parameter space (grid 2 in table \ref{t2}), we relaxed the parameter that sets the thickness of the pure helium envelope. We also never fixed the effective temperature and treated it as a free parameter within a reasonable range. We now come back to Fig. \ref{f1}, which actually shows quite an interesting result. On that graph, we show the helium abundance profile for our best fit model along with the profile calculated by Dehner \& Kawaler (1995). Two things match beautifully: 1) The profiles themselves (of course the base of the mixed He/C layer matches perfectly because we fixed that, but the base of the pure helium layer was eventually allowed to vary) and 2) The effective temperature. The latter was allowed to vary between 24100\ K and 30000\ K and settled on 29200\ K.

\section{Discussion}
\label{discussion}

This last result may be evidence that the calculations of time dependent diffusion calculations of Dehner \& Kawaler (1995), as well as others since then e.g. Althaus et al. (2005) describe the evolution of PG1159 stars well. However, we also need to look at what the core structure is. Does it agree with stellar evolution calculations? The answer there is not quite as positive. We find a central oxygen abundance of 0.60. With the standard reaction rates and treatment for convection, we expect the central abundance for a white dwarf of mass 0.550 $M_\odot$ to be between $\sim 0.8$ and $\sim 0.9$ (Althaus et al.~2010; Salaris et al.~1997). Perhaps more importantly, as our models are highly sensitive to that parameter (Bischoff-Kim \& Metcalfe 2011), the location of the edge of the homogeneous core is further inward than what stellar evolution calculations find.

In this paper, we took three different approaches using the pulsations to determine an effective temperature for WD\ J1929+4447: 1) Through a quick inspection of the pulsation spectrum, noting that low period modes were observed, 2) using average period spacing arguments, and 3) by performing asteroseismic fits of the period spectrum. All three point to an effective temperature more similar to EC\ 20058-5234's and inconsistent with the spectroscopic value of $\rm T_{eff}=24900 \pm 750$\ K. A well-calibrated high-S/N spectrum is urgently needed to settle the temperature issue.

\section{Conclusions and Future Work}
\label{conclusions}

We performed an asteroseismic analysis of WD\ J1929+4447, a DBV discovered in the field of view of the \emph{Kepler} satellite. We find strong evidence that the star is a hot DBV. Our models also support the time dependent diffusion calculations of Dehner \& Kawaler (1995), though more positive results from the study of other DBVs are needed to further support these results. Also, it would be worth approaching the problem from another angle. In section \ref{grids} we made the somewhat arbitrary choice of fixing the envelope parameters and allowing the core parameters to vary. We would most likely learn a lot by trying to fix the core parameters to those expected from stellar evolution (e.g. Althaus et al.~2010; Salaris et al.~1997) and allow the envelope parameters to freely vary over the entire reasonable range. Continuous observations with the \emph{Kepler} satellite may also reveal more modes, hidden in noise at the current S/N level, allowing 6 parameter fits.

The discovery of a hot DBV is significant, as it gives us another chance to study plasmon neutrino emission from white dwarfs. At 29200\ K, WD\ J1929+4447 should be radiating as much energy through the emission of photons as through the emission of plasmon neutrinos (Winget et al.~2004). The extra cooling due to neutrinos is therefore easily detectable if one could measure the cooling rate for the star with any significance. An observed period change in any of the modes can be a reliable estimator for the cooling rate, and only requires that at least one of the observed modes are phase-stable over a sufficient number of years to detect the variation. If \emph{Kepler} is permitted to maintain observations of WD J1929+4447 for a period of five years, the cooling rate should be determined to a sufficient precision to establish the respective contributions from thermal radiation and plasmon neutrinos.



\clearpage

\end{document}